\input{aipcheck}

\documentclass[
    ,final            
  ]
  {aipproc}

\layoutstyle{8x11double}

\begin{document}

\title{Neutrino induced coherent pion production}


\author{E. Hern\'andez}{
  address={Departamento de F\'{\i}sica Fundamental e IUFFyM, Universidad de
  Salamanca, E-37008 Salamanca, Spain},
}
\author{J. Nieves}{
  address={Instituto de F\'{\i}sica Corpuscular (IFIC), Centro Mixto
  CSIC-Universidad de Valencia, Institutos de Investigaci\'on de Paterna,
  Aptd. 22085, E-46071 Valencia, Spain},
}
\author{M. Valverde}{
  address={Research Center for Nuclear Physics (RCNP), Osaka University,
Ibaraki 567-0047, Japan},
}
\author{M.J. Vicente-Vacas}{
  address={Departamento de F\'{\i}sica Te\'orica e IFIC, Centro Mixto
  CSIC-Universidad de Valencia, Institutos de Investigaci\'on de Paterna,
  Aptd. 22085, E-46071 Valencia, Spain},
}

\begin{abstract}
We discuss different  parameterizations of the $C_5^A(q^2)$ $N\Delta$ axial form factor, 
 fitted to the 
old Argonne bubble chamber data for pion production by neutrinos, and we use coherent pion production
to test their low $q^2$ behavior.
We find  moderate effects that will be difficult to observe with the accuracy of present experiments.
We also discuss the use of the Rein-Sehgal model for low energy coherent pion production.
By comparison to a microscopic calculation, we show the weaknesses of some of the approximations
in that model that lead to very large   cross sections as well as to the wrong shapes for
differential ones. Finally we show that models based on the partial conservation of the axial current
hypothesis  are not fully reliable for differential
cross sections that depend on the angle formed by the pion and the incident neutrino.

\end{abstract}

\maketitle


\section{Introduction}
A proper
understanding of coherent pion production is very important in the analysis of
neutrino oscillation experiments since pion production is a source of 
background~\cite{AguilarArevalo:2007it}.  With pions being mainly produced through the excitation 
of nucleon resonances, coherent production 
can be used to extract information on axial form factors for the
nucleon-to-resonance transition.  

In coherent production the  nucleus remains in its ground
state and the reaction is controlled by the nucleus form factor. The nucleus form factor favors
small values of the nucleus momentum transfer squared $t$. Small $t$  values imply in this reaction small
$q^2$ (square of the lepton momentum transfer). For small $q^2$,
coherent pion production is dominated by
the divergence of the axial current
and it can thus be related to the pion-nucleus coherent scattering through the
partial conservation of the axial current (PCAC) hypothesis.

Experimental analyses of the coherent reaction rely on the Rein--Sehgal (RS) model~\cite{RS}
which is based on PCAC. In the RS model the pion-nucleus coherent cross section is written in
terms of the pion-nucleon elastic cross section by means of approximations that are valid for high neutrino
energies and  small $t$ and $q^2$ values. 
As  pointed out in Refs.~\cite{amaro,eli1}, those approximations are not reliable  for neutrino energies 
below/around 1 GeV and  light nuclei, like carbon or oxygen.

There are other approaches to coherent production that do not rely on PCAC but on microscopic models for pion 
production at the nucleon level~\cite{amaro,refnuc}. The dominant contribution to  the elementary amplitude at 
low energies is given  by the $\Delta$-pole mechanism ($\Delta$ excitation and its subsequent
decay into pion nucleon). Medium effects like the  modification of the $\Delta$ mass and width in
 the medium, final pion distortion, evaluated by solving the Klein-Gordon equation for a 
 pion-nucleus optical potential, as well as nonlocalities in the pion momentum, are very important
  and are taken into account in
 microscopic calculations. In the microscopic model of Ref.~\cite{eli2} background terms were 
included on top of the dominant $\Delta$-pole contribution.  The least known
ingredients of the model are the axial $N\Delta$ transition
form factors, of which $C_5^A$ gives the largest contribution (See Eq.~1 of Ref.~\cite{Schreiner:1973ka}
 for a form factor  decomposition of the $N\Delta$ weak current). Besides, within
the Adler model~\cite{adler}  used in Ref.~\cite{eli2}, $C_5^A$ determines all other axial
form factors. This
strongly suggested the readjustment of that form factor to the experimental
data, which the authors did by fitting the flux-averaged $\nu_\mu p\to \mu^- p\pi^+$ 
Argonne (ANL)~\cite{Barish:1978pj,Radecky:1981fn} $q^2$-differential cross section for pion-nucleon invariant
masses $W < 1.4$ GeV, for which the model should be appropriate. 
They found  $C_5^A(0)=0.867$ which is some 30\% smaller than the
value predicted by the  off-diagonal Goldberger-Treiman (GT) relation that
predicts $ C_5^A(0)=1.2$.  Background terms turn out to play a minor role in coherent production~\cite{amaro}
 where a reduced $C_5^A(0)$ value  gives rise to  smaller cross sections.
Coherent production is dominated by the axial current 
which in microscopic models is, in its turn, dominated by  the $C_5^A$ axial form factor. Coherent 
production could then
 be used to test the validity of different $C_5^A(q^2)$  dependences in the small $q^2$ region. 

In this contribution  we will concentrate on the two issues mentioned before: first, we will try to see
 how sensitive  coherent production is to different $C_5^A(q^2)$ parameterizations  proposed in the
literature. Second, we will present results that show how and why the RS model fails for low energy
pion coherent production on light nuclei.

\section{$C_5^A(q^2)$ in coherent production }
As mentioned before, in Ref.~\cite{eli2} the authors made a fit of $C_5^A(q^2)$ to ANL data. 
For $C_{5}^A$ they took the $q^2$ dependence of Ref.~\cite{Pa04} 
\begin{equation}
C_5^A(q^2) = \frac{C_5^A(0)}{(1-q^2/M^2_{A\Delta})^2}\times
\frac{1}{1-\frac{q^2}{3M_{A\Delta}^2}}, 
\label{eq:c5a}
\end{equation}
and from the fit they obtained
\begin{equation}
C_5^A(0) = 0.867 \pm 0.075,\,  M_{A\Delta}=0.985\pm 0.082\,{\rm
  GeV}.
  \label{eq:parnew}
\end{equation}
 This fitted axial mass in the weak $N\Delta$ vertex
is in good agreement with the estimates of about 0.95 GeV and 0.84 GeV
given in the original ANL reference~\cite{Radecky:1981fn} and in the work of
Ref.~\cite{Pa05}. On the other hand, a  correction of the
order of 30\% to the off-diagonal GT relation value was found
for $C_5^A(0)$. 
As shown in Ref.~\cite{eli2} the agreement with  ANL total cross sections  improved
 with the fitted  values
for $C_5^A(0)$ and $M_{A\Delta}$. On the other
hand it is also shown that Brookhaven (BNL) bubble chamber data~\cite{bnl}  and ANL data are not fully compatible
and that BNL data alone would favor a $C_5^A(0)\approx 1.2$ value as
given by the off-diagonal GT relation.

\begin{figure}
 \includegraphics[width=7.5cm]{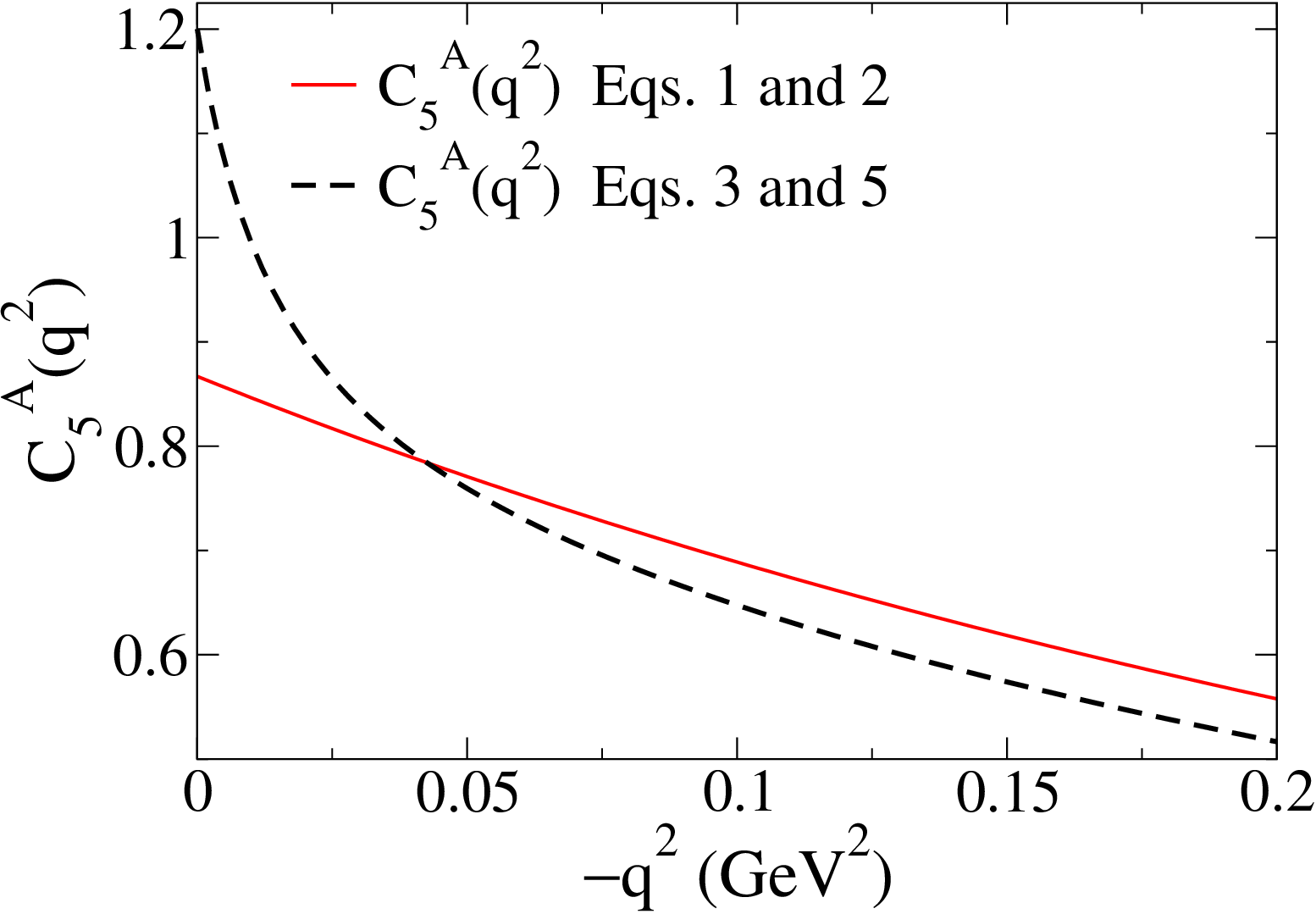}
\caption{Different $C_5^A(q^2)$ parameterizations fitted to ANL data.}
\label{fig:c5a}
\end{figure}

A different approach has been followed by Leitner {\it et al.} in
Ref.~\cite{leitner09}. There, the authors use a different
parameterization for $C_5^A(q^2)$
\begin{equation}
C_5^A(q^2) = 
 {1.2\times(1-\frac{{ a} q^2}{{ b}-q^2})}/{\left(1-q^2/{
 M_{A\Delta}}^2\right)^2},
 \label{eq:c5amosel}
\end{equation}
in which $C_5^A(0)$ is kept to its off-diagonal GT relation value
$C_5^A(0)=1.2$, while $a,\,b$ and $M_{A\Delta}$ are treated as free
parameters. One can accommodate a larger $C_5^A(0)$ value by changing
the $q^2$ dependence. In fact very small $-q^2$ values are not very
relevant due to phase space and what is actually important is the
$C_5^A(q^2)$ value in the region around $-q^2\approx0.1\,$GeV$^2$.
When fitting the ANL data with the $\Delta$-pole term alone they
got~\cite{leitner09}
\begin{equation}
a=-0.25,\  b=0.04\,{\rm GeV}^2,\  M_{A\Delta}=0.95\,{\rm GeV}.
\label{eq:parmosel}
\end{equation}
 But background terms are important at the nucleon level and they should be included 
 in the calculation. A new fit including background terms 
leads to 
\begin{eqnarray}
&&\hspace*{-.6cm}a=-0.3861\pm0.198,\  b=0.01536\pm0.0310\,{\rm GeV}^2,\nonumber\\
&& M_{A\Delta}=0.952\pm0.205\,{\rm GeV}.
\label{eq:parfullmosel}
\end{eqnarray}
Both the fit in Eqs.~\ref{eq:c5a} and \ref{eq:parnew} and the one in  Eqs.~\ref{eq:c5amosel} and 
\ref{eq:parfullmosel} give a good description of
ANL data even though they use quite different values for $C_5^A(0)$. 
The different $q^2$ dependence compensates for the initial difference at
$q^2=0$. Note, however, the large errors in Eq.~\ref{eq:parfullmosel} that point at large
correlations between the different parameters.
In Fig.~\ref{fig:c5a} we compare the two form factor parameterizations. The main
differences are in the $-q^2<0.025\,$GeV region. The corresponding  axial radii are
$R_A^2=0.56^{+0.10}_{-0.09}\,$fm$^2$ for the parameterization given by  
Eqs.~\ref{eq:c5a} and \ref{eq:parnew}, and $R_A^2=6.4^{+0.9}_{-8.4}\,$fm$^2$ for the 
one
 given by  Eqs.~\ref{eq:c5amosel}  and \ref{eq:parfullmosel}. The 
large negative error in the latter case is a reflection of the large statistical errors in
Eq.~\ref{eq:parfullmosel}. Both things point to the fact that Eq.~\ref{eq:c5amosel}
might  not be a good parameterization.

In the following we will try to see if coherent pion production can give us extra information on the
validity of the two different $C_5^A(q^2)$ discussed above. Here we shall use the coherent
production model of Ref.~\cite{amaro}.

\begin{figure}
\includegraphics[width=7.25cm]{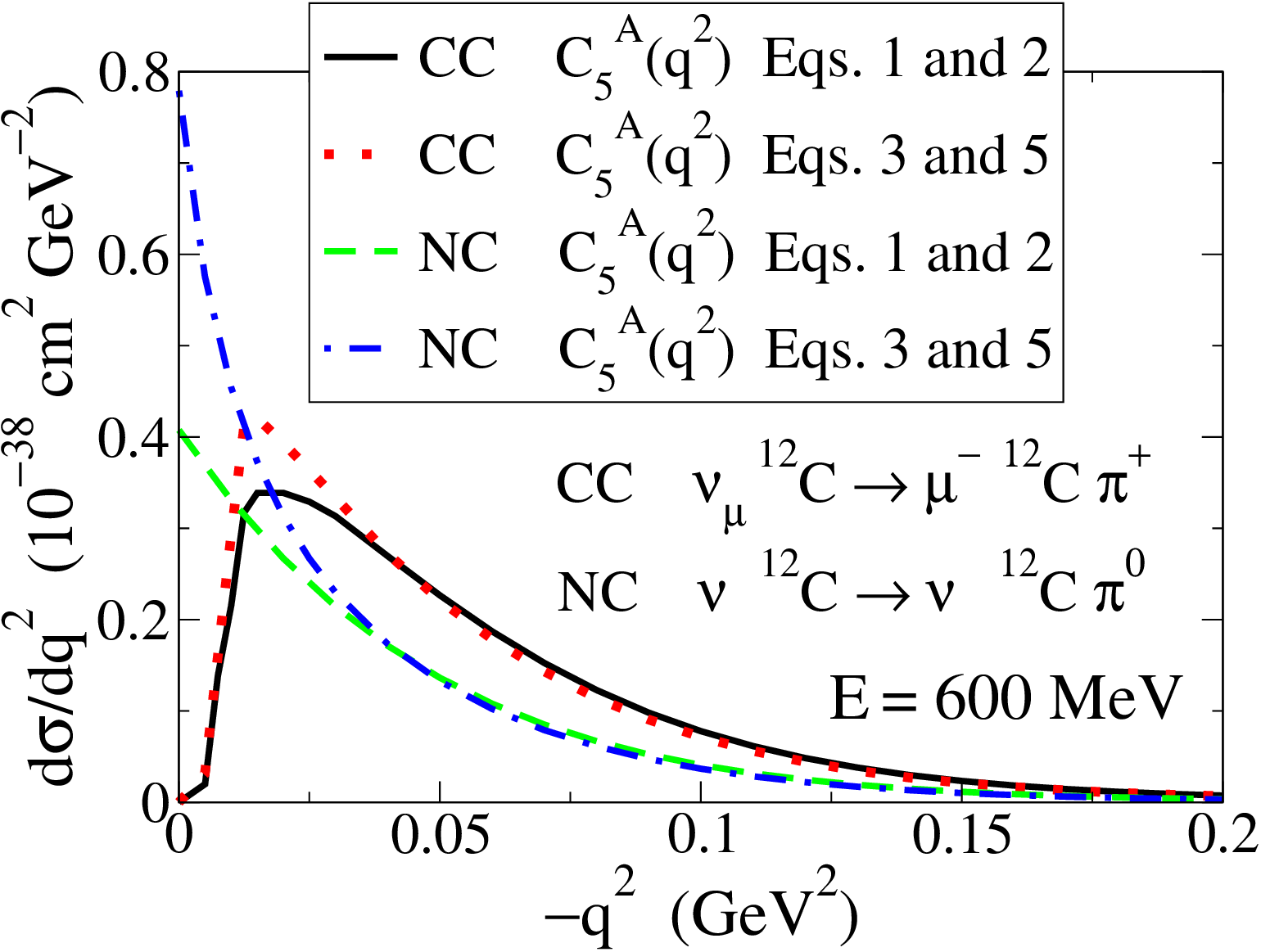}
\caption{$\frac{d\sigma}{dq^2}$ differential cross sections for CC and NC 
neutrino induced one-pion production  on carbon obtained with different $C_5^A(q^2)$ parameterizations. The incident neutrino energy is $E=600\,$MeV.}
\label{fig:coh_q2}
\end{figure}

In Fig.~\ref{fig:coh_q2} we show the $q^2$ differential cross section
for charged current (CC) and neutral current (NC) coherent production
on carbon and for an incident neutrino energy $E=600\,$MeV. In both cases one can see
differences for low $-q^2$ due to the different $C_5^A(q^2)$ form
factor used. These effects are more relevant in the NC case where
smaller $-q^2$ values are not suppressed by phase space.  As
$\frac{d\sigma}{dq^2}$ can not be measured for the NC channel  one
has to rely on total cross sections. There, the change is a mere
3.4\% for the CC channel, while for the NC process, where lower $-q^2$
can be reached, the change is 16.6\%.

\begin{figure}
\includegraphics[width=7.25cm]{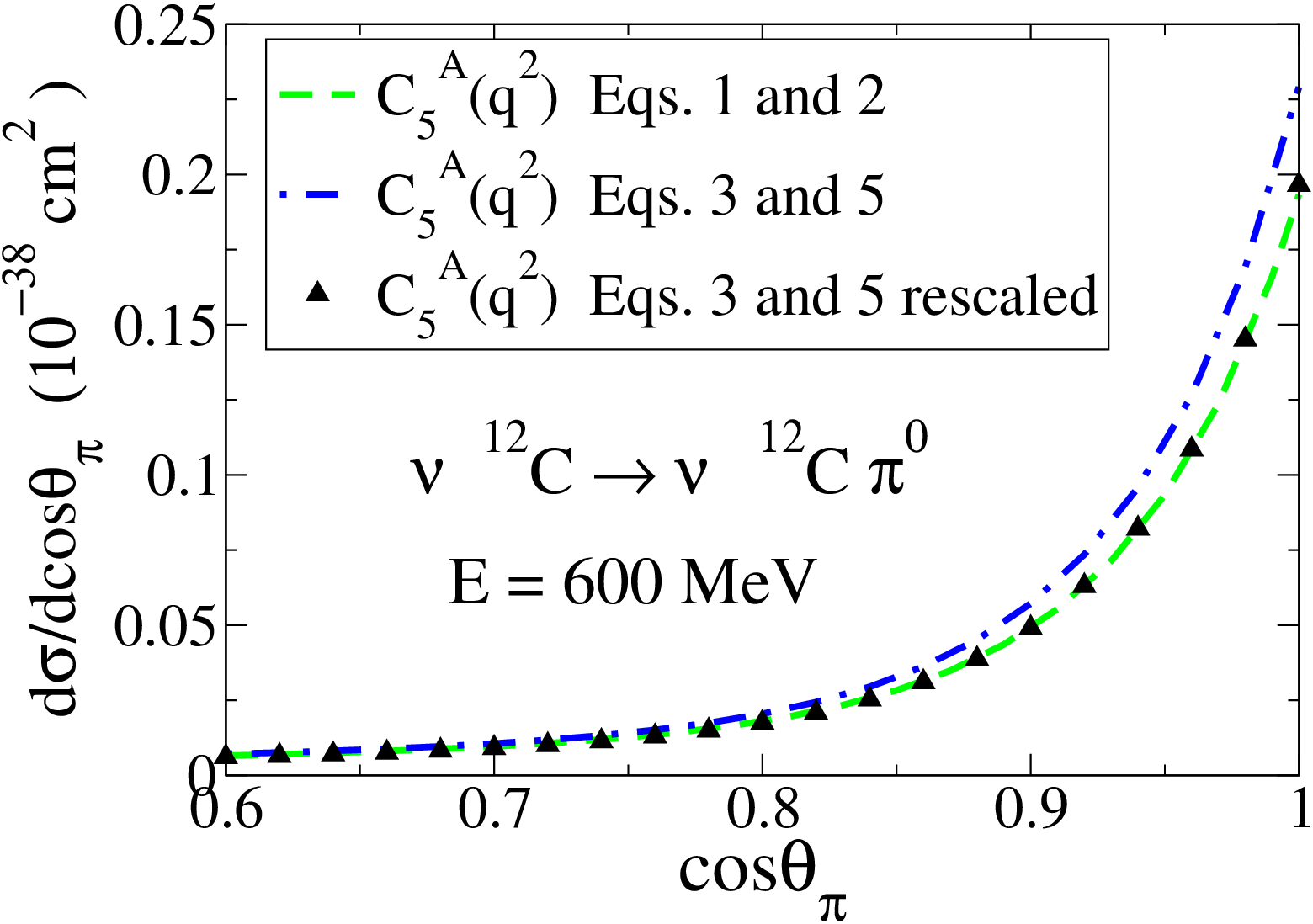}
\caption{$\frac{d\sigma}{d\cos\theta_\pi}$ differential cross section for the
$\nu\ ^{12}C \to \nu\ ^{12}C\ \pi^0$ reaction obtained with different $C_5^A(q^2)$ parameterizations. The incident neutrino energy is $E=600\,$MeV}
\label{fig:coh_cos}
\end{figure}

In Fig.~\ref{fig:coh_cos} we show the NC differential cross section with
respect to $\cos\theta_\pi$, being  $\theta_\pi$ the angle formed by
the pion and the incident neutrino. In this case there is no
difference in shape as can be appreciated by comparing the solid line
and the up triangle line. A similar result is obtained for the CC case
and for the NC/CC differential cross
section with respect to the pion kinetic energy.

\begin{figure}
\includegraphics[width=7.25cm]{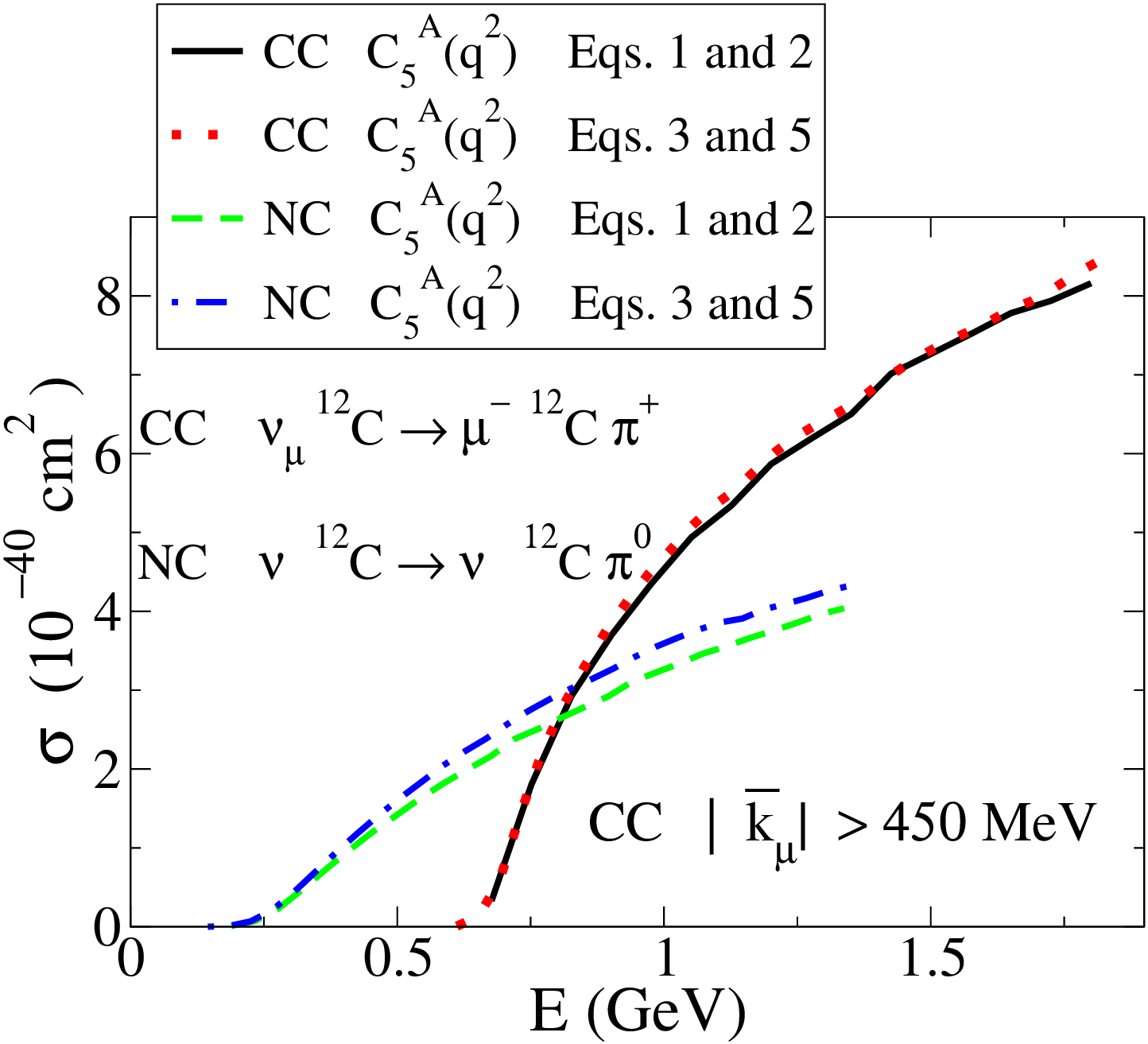}
\caption{Total  cross sections, as a function of the incident neutrino energy, for CC and NC neutrino induced one-pion production on carbon obtained with different $C_5^A(q^2)$ parameterizations. For the CC case a cut $|\vec k_\mu|>450\,$ MeV on the final muon momentum has been applied.}
\label{fig:coh_tot}
\end{figure}

Finally in Fig.~\ref{fig:coh_tot} we show total cross sections. For
the CC case  a cut $|\vec k_\mu|>450\,$MeV on the final
muon momentum, as the one imposed by the K2K
Collaboration~\cite{Hasegawa:2005td}, has been applied. The changes are more
significant for the NC case.
 In Table~\ref{tab:res} we show the CC results convolved
with the K2K flux and the NC ones convolved with the MiniBooNE
flux. Details of the convolution can be found in
Ref.~\cite{amaro}. We also show for comparison the experimental
data. The change in the parameterization for $C_5^A(q^2)$ amounts to
an small increase of 3\% for the CC/K2K case. Both results are below the
present experimental upper bound. For the
NC/MiniBooNE case there is an increase of 10.7\%, with the results of both
parameterizations being a factor of 2 smaller than present experimental
data~\cite{anderson}.

Coherent pion productions shows a moderate dependence on different $C_5^A(q^2)$ 
parameterizations fitted to the ANL data. With present experimental accuracy it 
can not be used to constraint the form factor small $q^2$ behavior.

\begin{table}
    \begin{tabular}{lccccc}\hline
Reaction                 & \hspace*{.5cm}Experiment\hspace*{.5cm} &$C_5^A(q^2)$&
$\bar \sigma$[$10^{-40}$cm$^2$]& $\sigma_{\rm exp}$[$10^{-40}$cm$^2$] 
\\\hline
CC\phantom{*} $\nu_\mu + ^{12}$C    & K2K        & Eqs.~\ref{eq:c5a} and \ref{eq:parnew} & 4.68   
       \vspace{-.3cm} \\
&&&& $\bigg\}\ <7.7 $~\cite{Hasegawa:2005td}\vspace{-.3cm}\\
CC\phantom{*} $\nu_\mu + ^{12}$C    & K2K     & Eqs.~\ref{eq:c5amosel} and
\ref{eq:parfullmosel}   &   4.82    
       \\\hline
NC\phantom{*} $\nu_{{\mu}} + ^{12}$C    & MiniBooNE & Eqs.~\ref{eq:c5a} and 
\ref{eq:parnew}  & 1.97
\vspace{-.3cm} \\
 &   &&  & $\bigg\}\ 4.54\pm0.04\pm0.71$~\cite{anderson}
\vspace{-.3cm}\\
NC\phantom{*} $\nu_{{\mu}} + ^{12}$C    & MiniBooNE & Eqs.~\ref{eq:c5amosel} 
and \ref{eq:parfullmosel} & 2.18 \vspace*{.2cm} \\
\hline
    \end{tabular}
  \caption{Total cross sections for CC and NC neutrino induced one-pion production on carbon convolved with the K2K and MiniBooNE fluxes. Different $C_5^A(q^2)$ parameterizations have been used. Experimental results are also shown.}
\label{tab:res} 
\end{table}

\section{Rein-Sehgal model for low energy coherent production}
The RS model is based on PCAC, so it is worth motivating how one constructs a general PCAC based model.
Let us look, for simplicity, at the NC process
\begin{eqnarray}
  \nu_l (k(E,\vec k)) + {\cal N}_{gs} \to \nu_l (k^\prime(E',\vec k')) + {\cal N}_{gs} +
  \pi^0(k_\pi)\, .
\end{eqnarray}
Defining $q=k-k^\prime$, $y=q^0/E$, $t=(q-k_\pi)^2$, taking $\vec q$
along the positive $Z$ axis and $\vec k\times\vec k'$ along the positive $Y$axis,
and calling $\phi_{\pi q}$ to the  pion azimuthal angle in the $XYZ$ frame,
Lorentz invariance allows us to write 
{\small\begin{eqnarray}
&&\hspace*{-1cm}\frac{d\sigma}{dq^2dy\,dt\, d\phi_{\pi  q}} = \frac{G^2}{16\pi^2} 
\frac{-q^2E\, \kappa}{|\vec{q }\,|^2}\nonumber\\
&&\hspace{-.5cm}\times\left(\frac{u^2}{2\pi}\frac{d
  \sigma_L}{dt}+\frac{v^2}{2\pi}\frac{d\sigma_R}{dt}
+2\,\frac{u v}{2\pi}\frac{d\sigma_S}{dt}+\frac{d{\cal A}}{dt\, d\phi_{\pi
q}}\right), 
\label{eq:dsig1}
\end{eqnarray}}
where $G$ is the Fermi decay constant, $\kappa=q^0+\frac{q^2}{2{\cal M}}$, with ${\cal M}$ the
nucleus mass, and 
$u,v=\frac{E+E'\pm |\vec{q}\,|}{2E}$. Besides,
$\sigma_{R,L,S}$ stand for cross sections for right, left and scalar
polarized intermediate vector mesons. ${\cal A}$ is not a proper cross section and it 
contains all the dependence on $\phi_{\pi q}$.  As shown in Ref.~\cite{eli1}, Eq.~\ref{eq:dsig1} should be the starting point
to evaluate cross sections with respect to $\theta_\pi$. PCAC based models take instead as a starting point  
{\small \begin{eqnarray}
\frac{d\sigma}{dq^2dy\,dt} = \frac{G^2}{16\pi^2} 
\frac{-q^2E\, \kappa}{|\vec{q }\,|^2}
\left({u^2}\frac{d
  \sigma_L}{dt}+{v^2}\frac{d\sigma_R}{dt}
+2\,{u v}\frac{d\sigma_S}{dt}\right), 
\label{eq:dsig2}
\end{eqnarray}}
which is obtained from Eq.~\ref{eq:dsig1} after integration on
$\phi_{\pi q}$, 
and they further assume 
{\small\begin{equation}
\frac{d\sigma}{dq^2\,dy\,dt\,d\phi_{\pi q}}=\frac{1}{2\pi}\frac{d\sigma}{dq^2dy\,dt}.
\label{eq:phipiq}
\end{equation}}
The latter is 
incorrect for $q^2\neq 0$ and it will have consequences when evaluating cross sections with respect to
variables that depend on $\theta_\pi$.

For $q^2=0$ only $\sigma_S$ contributes and one finds that $\left.q^2\frac{d\sigma_S}{dt}\right|_{q^2=0}$
is given as the modulus square of the hadronic matrix element of the divergence of the weak current.
Since the vector NC current is conserved one is left with the divergence of the
axial current that can be related, through PCAC, to the pion-nucleus elastic cross section
{\small\begin{eqnarray}
\left.q^2\frac{d\sigma_S}{dt}\right|_{q^2=0}=-4\frac{k^0_\pi}{\kappa}f^2_\pi\left.
\frac{d\sigma(\pi^0{\cal N}_{gs}\to \pi^0{\cal N}_{gs})}{dt}\right|_{q^2=0}.
\end{eqnarray}}
with $f_\pi=92.4\,$MeV. Neglecting the nucleus recoil ($k^0_\pi=q^0$) and including a form
factor $G_A=1/(1-q^2/m_A^2)$ for $q^2\neq 0$ ,
one arrives at
 {\small\begin{eqnarray}
\frac{d\sigma}{dq^2\,dy\,dt}=\frac{G_F^2f^2_\pi}{2\pi^2}\frac{Euv}{|\vec q\,|}G_A^2
\frac{d\sigma(\pi^0{\cal N}_{gs}\to\pi^0{\cal N}_{gs} )}{dt}.
\end{eqnarray}}
This is the Berger-Sehgal model for $\pi^0$ coherent  production~\cite{BS}. 
In the  RS model they further approximate
{\small \begin{eqnarray}
&&\frac{Euv}{|\vec q\,|}\to\frac{1-y}{y}\ \ {({\rm exact\ for\ }\  q^2=0)},\\ 
&&\hspace{-.75cm}\frac{d\sigma(\pi^0 {\cal N}_{gs} 
\to \pi^0 {\cal N}_{gs})}{dt}\to|F_A(t)|^2\ F_{\rm abs}
\left.\frac{d\sigma(\pi^0 { N} 
)}{dt}\right|_{t=0},
\end{eqnarray}}
with $F_A(t)$ the nucleus form factor, $F_{\rm abs}$ a $t-$independent eikonal absorption factor that takes into
account the distortion of the final pion, and $\left.\frac{d\sigma(\pi^0 { N} 
)}{dt}\right|_{t=0}$  the elastic pion-nucleon differential cross section at $t=0$.

It is worth  modifying the RS model by improving  some of its approximations~\cite{eli1}. The $t=0$ 
approximation can be eliminated altogether by substituting
\begin{equation}
\left.\frac{d\sigma(\pi^0 { N} 
)}{dt}\right|_{t=0} \longrightarrow \frac{d\sigma_{nsp}(\pi^0 { N} 
)}{dt},
\label{eq:nsp}
\end{equation}
where $nsp$ stands for the non-spin-flip part of the pion-nucleon elastic cross section. Besides,
the pion distortion can be improved, still in an eikonal approach, with the replacement
{\normalsize \begin{eqnarray}
|F_A(t)|^2\,F_{abs}\longrightarrow \left|\int d^3\vec r\ e^{i(\vec q-\vec k_\pi)\cdot \vec r}
\, \rho(\vec r) \Gamma(b,z)\right|^2,\nonumber\\
\Gamma(b,z)=\exp\left(-\frac12\sigma_{inel}\int_z^\infty dz' 
 \rho(\sqrt{b^2+z'^2})\right).
 \label{eq:distort}
\end{eqnarray}}
Finally one can use  $\frac{Euv}{|\vec q\,|}$ instead of $\frac{1-y}{y}$. 

\begin{figure}
 \includegraphics[width=7.5cm]{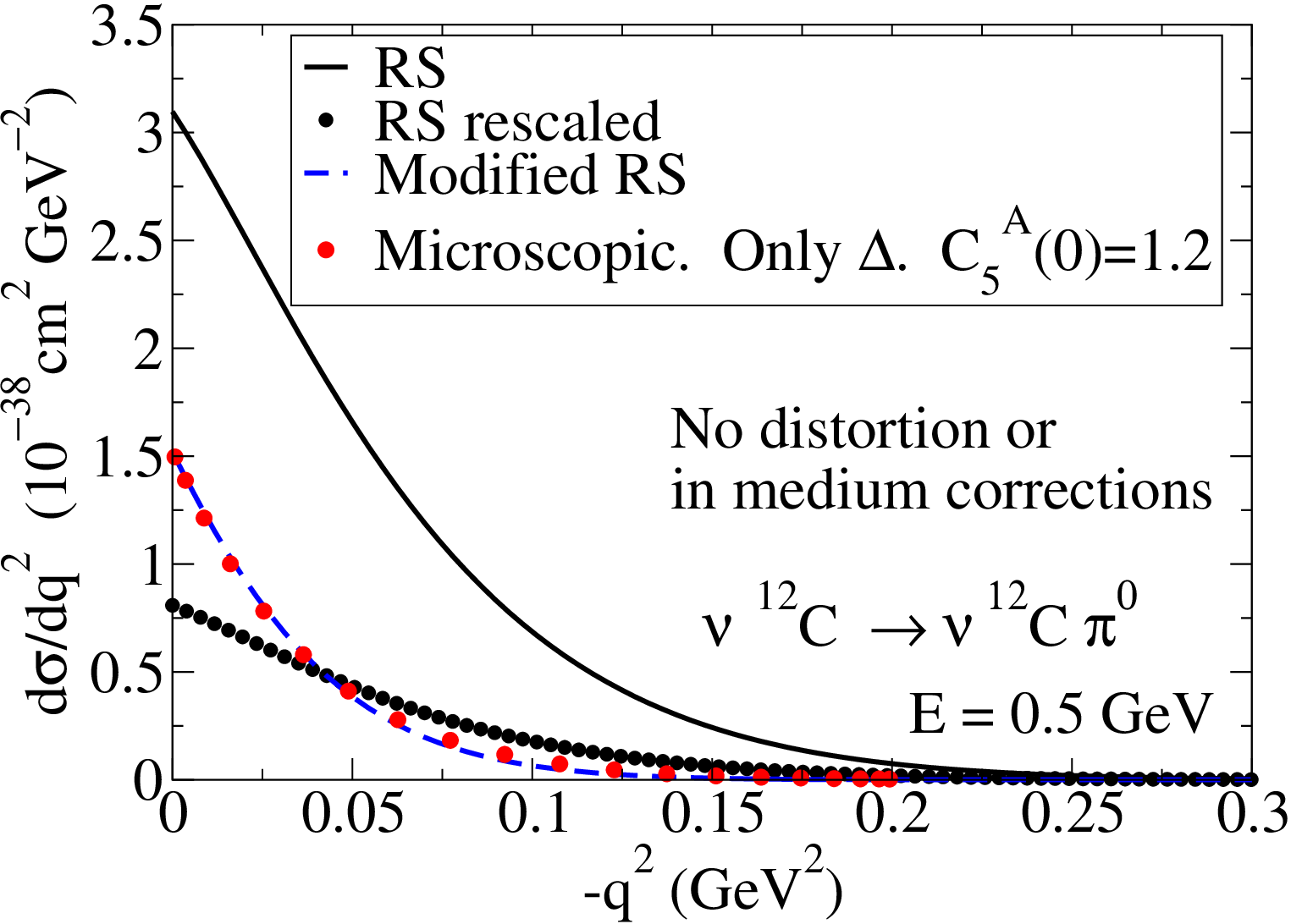}
  \caption{$\frac{d\sigma}{dq^2}$ differential cross sections for the
  $\nu\ ^{12}C\to \nu\ ^{12}C\ \pi^0$ reaction obtained within different models. The incident neutrino energy is
  $E=0.5\,$GeV. No distortion or in medium corrections were considered in the calculations.}
  \label{fig:dsigmadq2}
\end{figure}

In Fig.~\ref{fig:dsigmadq2} we evaluate $\frac{d\sigma}{dq^2}$ for NC coherent production on 
carbon at  $E=0.5\,$GeV. In order to check the $t=0$ approximation of the RS model, no distortion is included in the
calculation. We compare  with the modified RS model where the $t=0$ approximation has
been removed. We also compare with the microscopic calculation of Ref.~\cite{amaro} that we
consider a good model for coherent production at these energies. For simplicity, in the microscopic
 calculation we have only kept the dominant  $\Delta$ contribution and, in order
 to make the comparison meaningful, we  have
 taken $C_5^A(0)=1.2$ (to fix 
normalization) and we have not included any in medium correction for the $\Delta$ or any pion 
distortion. We see that the modified RS model compares very well with the microscopic calculation,
 whereas the original RS model fails to reproduce both the size  and the shape of the differential 
 cross section.  

\begin{figure}
 \includegraphics[width=7.5cm]{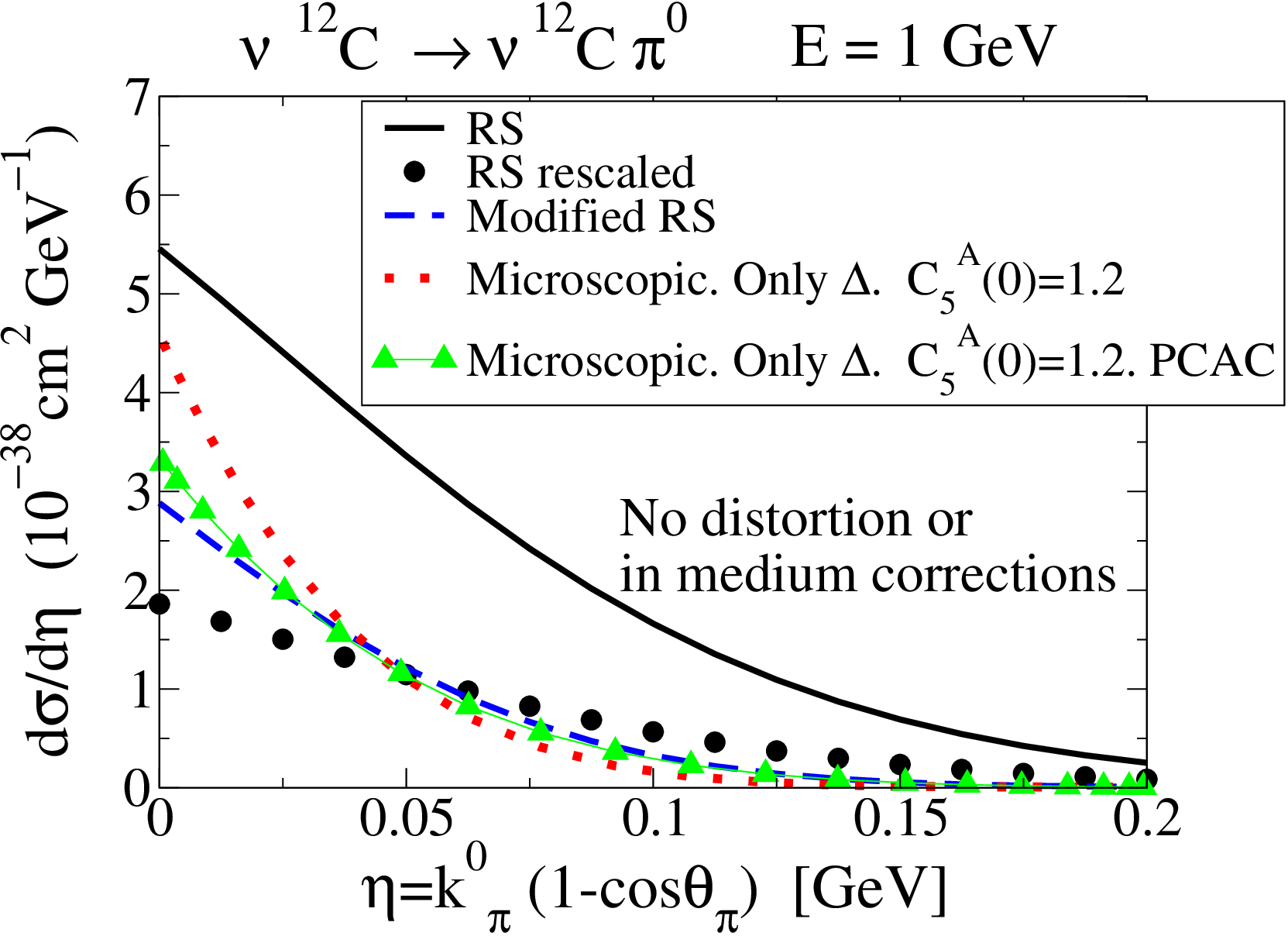}
  \caption{$\frac{d\sigma}{d\eta}$ differential cross sections for the
  $\nu\ ^{12}C\to \nu\ ^{12}C\ \pi^0$ reaction obtained within different models. The incident neutrino energy is
  $E=1\,$GeV. No distortion or in medium corrections were considered in the calculations.}
  \label{fig:dsigmadeta}
\end{figure}

In the same conditions as before, we show in Fig.~\ref{fig:dsigmadeta}  the $\frac{d\sigma}{d\eta}$ differential cross section
on carbon at $E=1\,$GeV. Once more, we see the size
and shape of the RS calculation is very different from the microscopic one, showing the inadequacy
of the $t=0$ approximation. In this case the
modified RS is also in disagreement with the microscopic calculation. This is mainly due  to the
approximation in any PCAC based model encoded in Eq.~\ref{eq:phipiq} and, to a lesser extent, to the
neglect by PCAC models of the $\sigma_{R,L}$ contributions. This assertion can be checked by eliminating from 
the microscopic calculation the non-PCAC contributions, i.e. taking only $\sigma_S$ and assuming 
Eq.~\ref{eq:phipiq}. The results are shown in Fig.~\ref{fig:dsigmadeta} with
up triangles. We now see the agreement with the modified RS calculation
is good. This shows  in fact that PCAC based models are not fully reliable to give differential cross sections
with respect to variables that depend on $\theta_\pi$. 

\begin{figure}
 \includegraphics[width=7.5cm]{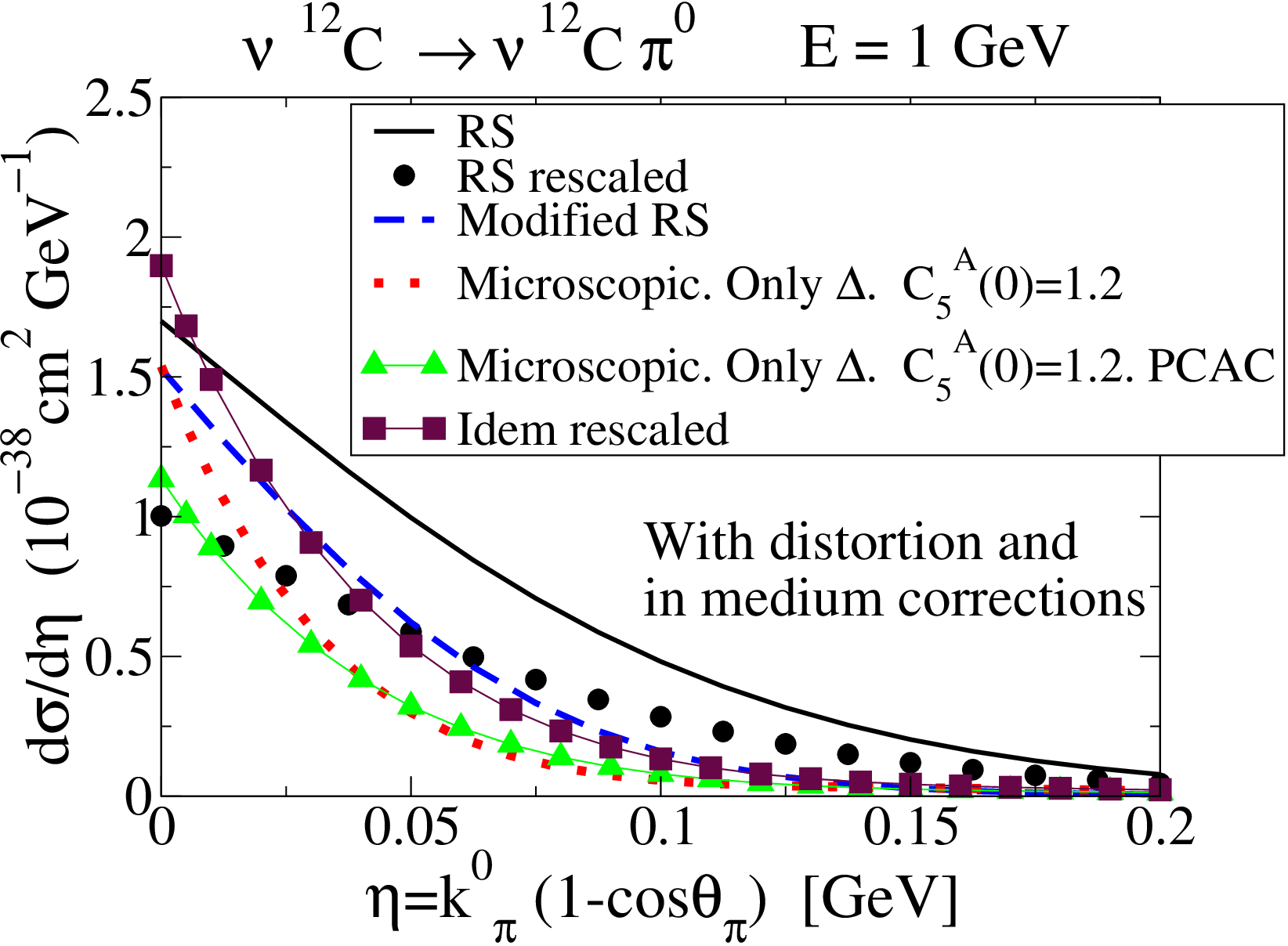}
  \caption{Same as Fig.~\ref{fig:dsigmadeta}, but now with pion distortion and
  in medium corrections taken into account.}
  \label{fig:dsigmadeta2}
\end{figure}

Last, in Fig.~\ref{fig:dsigmadeta2} we present 
results  for $\frac{d\sigma}{d\eta}$ but now  with distortion and in medium corrections. Again, the sizes and shapes of the RS model
predictions are very different from the microscopic  ones. The same is true for the 
modified RS  calculation, even when compared to the microscopic one with only PCAC
contributions. While the shapes of the differential cross sections of the two latter calculations are similar, the integrated cross sections differ by a factor $\approx
1.7$.  This hints at the inadequacy of eikonal factors to account for the final
pion distortion or  in medium modifications of the $\pi N$ cross section. We see again
that the approximations in PCAC models give rise to the wrong distributions for variables depending on $\theta_\pi$.

As shown, the $t=0$ approximation and the eikonal distortion used in the RS model are not adequate for
low energy coherent production on light nuclei. Besides, and due to the assumption in Eq.~\ref{eq:phipiq}, any PCAC model 
would predict the wrong shape for differential cross sections with respect to variables that depend 
on $\theta_\pi$.




\begin{theacknowledgments}
  This research was supported by DGI and FEDER funds, under contracts
  FIS2008-01143/FIS, FIS2006-03438, FPA2007-65748, CSD2007-00042, by Junta de
  Castilla y Le\'on under contracts SA016A07 and GR12, and by the EU 
  HadronPhysics2 project. M.V. wishes to acknowledge a postdoctoral fellowship form
  the Japan Society for the Promotion of Science.

\end{theacknowledgments}

\end{document}